\documentstyle[12pt]{article}

\def\beq{\begin{equation}}
\def\eeq{\end{equation}}
\def\bea{\begin{eqnarray}}
\def\eea{\end{eqnarray}}
\def\bq{\begin{quote}}
\def\eq{\end{quote}}

\def\CQG{{\it Class.Quantum Grav.} }

\begin{document}

\begin{titlepage}
\title{
DECOHERENCE DUE TO MASSIVE VECTOR FIELDS WITH GLOBAL SYMMETRIES}
\author{O. Bertolami\footnote{On leave from Departamento de F\'{\i}sica,
Instituto Superior T\'ecnico,
Av. Rovisco Pais, 1096 Lisboa, Portugal; also at Theory Division, CERN,
CH - 1211 Geneva 23,  Switzerland; e-mail:
Orfeu@vxcern.cern.ch and Bertolami@toux40.to.infn.it}}
INFN - Sezione Torino, Via Pietro Giuria 1, I-10125 Turin, Italy
%
%%%\and
and
\author{P.V. Moniz\footnote{Work supported by a Human Capital and Mobility
Fellowship from the European
Union (Contract ERBCHBICT930781); e-mail: prlvm10@amtp.cam.ac.uk}}
University of Cambridge, DAMTP, Cambridge, CB3 9EW, UK
%%%\endauthor

\begin{abstract}
Retrieval of classical behaviour in quantum cosmology is usually discussed in
the framework of minisuperspace models in the presence of scalar fields
together with the inhomogeneous modes of gravitational or
scalar fields. In this work we propose alternatively a  model
where the scalar field is
replaced by a massive vector field with global $U(1)$ or $SO(3)$ internal
symmetries.
\end{abstract}

\end{titlepage}

\medskip

The emergence of the classical properties from the quantum mechanics formalism
remains still essentially an open problem. Some progress has, however,
been achieved
through the so-called decoherence approach.  On
fairly general grounds, decoherence can be regarded as a procedure
where one considers the
system under study to be part of a more complex world and which interacts with
other subsystems, usually referred to as ``environment". This interaction leads
to the vanishing of the off-diagonal terms of the density matrix of the
system suppressing in this way quantum interference.

These ideas have been recently very much discussed in the
context of minisuperspace models in quantum cosmology (see \cite{OBPM} and
references therein).
In the literature concerning the emergence of classical behaviour from
quantum cosmological minisuperspace models it is most usual to consider
 as environment the inhomogeneous modes either of the
gravitational or of the scalar fields.
We propose alternatively a model where the self-interacting scalar field is
replaced by a massive vector field with U(1) or SO(3) global internal
symmetries
[1,3,4].
Preliminary work on the  system
with $SO(3)$ non-Abelian global symmetry, whose classical cosmology has been
studied in Ref. \cite {PP},
shows that
the ingredients necessary for the process of decoherence to take place are
present \cite{SS}.The system with $U(1)$ global symmetry has been analysed in
Ref. \cite {SS1}. Notice that the presence of a mass term
is an essential feature as this breaks the
conformal symmetry of the spin-1 field action which leads to a Wheeler-DeWitt
equation where gravitational and matter degrees of freedom decouple
\cite {LLL}.

In this contribution, we overview the results of Ref. [1] on the
decoherence process
of homogeneous and isotropic metrics in the presence of massive vector
fields with $U(1)$ and $SO(3)$ global symmetries.
The main idea is that role of the environment is played by the
inhomogeneous modes of the vector field. In the
$U(1)$ case the corresponding
minisuperspace
is one-dimensional as it has only a
single ``classical'' degree of freedom as physical observable,
the scale factor.  The
minisuperspace for the non-Abelian model
is two-dimensional due to the specific
Ansatz for the homogeneous modes of massive the spin-1 field [2].

The action of our model consists of a Proca field coupled with gravity
\cite{PP}:
\beq
S = \int d^4x \sqrt{-g}~~\left[ {1\over 2k^2} R + {1\over
8e^2}
{\rm Tr}(\mbox{\boldmath $F$}^{(a)}_{\mu\nu}\mbox{\boldmath
$F$}^{(a)\mu\nu}) +
 {1\over 2}~m^2
{\rm Tr}(\mbox{\boldmath $A$}^{(a)}_{\mu} \mbox{\boldmath
 $A$}^{(a)\mu})\right]
-{1\over k^2} \int_{\partial M} d^3x \sqrt{h} K ,
\label{1}
\eeq
where $k^2 = 8\pi M^{-2}_P$, $M_P$ being the Planck mass, $e$ is a gauge
coupling constant, $m$ the mass of the Proca field and
with $h_{ij} (i,j = 1,2,3)$ being the induced metric on the three-dimensional
boundary $\partial M$ of $M$, $h = \det (h_{ij})$ and $K = K^{\mu}_\mu$ is the
trace of the second fundamental form on $\partial M$.

In quantum cosmology one is concerned with spatially compact topologies
 and we shall consider
the Friedmann-Robertson-Walker (FRW)
ansatz for the  ${\bf R}\times S^3$  geometry
\beq
ds^2 = \sigma^2 a^2(\eta) \left[ -N(\eta)^2 d\eta^2 + \sum^3_{i=1}
\omega^i\omega^i\right]~,
\label{2}
\eeq
 where
 $\sigma^2 = 2/3\pi$, $\eta$  is the conformal time,
$N(\eta)$ and $a(\eta)$ being the lapse function
and the scale factor, respectively and $\omega^i$ are left-invariant
one-forms in $SU(2)\simeq$ $S^3$.

The massive vector field

\begin{equation}
\mbox{\boldmath $A$} = A_m^{ab}\omega_{s}^m {\cal T}_{ab} =
 A_m^{ab} \sigma^m_i \omega^i {\cal T}_{ab},
\label{vec}
\end{equation}
where $\omega_{s}^m$ denote the one-forms in a spherical basis with
$m=0,\pm 1$, $\sigma^m_i$ denotes a 3 x 3 matrix and
${\cal T}_{ab}$ are the $SO(3)$ group generators,
is expanded in spin-1 hyperspherical harmonics as \cite{OBPM}:
\bea
A_0(\eta,x^j) &=& \sum_{JMN} \alpha^{abJM}_{\phantom{^{JM}} N}(\eta) {\cal
D}^{J{\phantom J}M}_{{\phantom J}N} (g) {\cal T}_{ab}
\label{11a} \\
A_i(\eta,x^j) &=& \sum_{LJNM} \beta^{abMN}_{LJ} (\eta)
{}~Y^{1LJ}_{mNM}(g) \sigma^m_i~{\cal T}_{ab} \nonumber \\
 & = & {1 \over 2}\left[1 + \sqrt{{{2\overline{\alpha}}
\over
{3\pi}}} \chi(\eta)\right]\epsilon_{aib}{\cal T}_{ab}
+ \sum_{L'J'N'M'} \beta^{abM'N'}_{L'J'} (\eta)
{}~Y^{1L'J'}_{mN'M'}(g) \sigma^m_i~{\cal T}_{ab},
\label{11b}
\eea
where $\overline{\alpha} = e^2/4\pi$ and $\chi(\eta)$ a
 time-dependent scalar function.
$A_0$ is a scalar on each fixed time
hypersurface, such that it can be expanded in scalar harmonics
${\cal D}^{J{\phantom J}M}_{{\phantom J}N}(g)$.
The expansion of $A_i$ is performed in terms of the spin-1 spinor
hyperspherical harmonics, $Y^{1~LJ}_{m~MN}(g)$. Longitudinal and the
transversal harmonics correspond to $L = J$ and $L-J = \pm 1$, respectively.
The rhs of eqs.
(\ref{11a}) and (\ref{11b})
correspond to a decomposition in homogeneous and
inhomogeneous modes. For this decomposition we have used
the Ansatz for the homogeneous modes
of the vector field that is compatible with the FRW geometry as discussed in
Ref. \cite{PP}.

 From action (\ref{1})
one can work out the effective Hamiltonian density obtained from the
substitution of the
expansions (\ref{2})-(\ref{11b}). To second order in the coefficients of the
expansions and in all orders in $a$, one obtains
the following effective Hamiltonian density for the system with SO(3) global
symmetry (for the Abelian case one drops the last four terms) \cite{OBPM}:
\bea
{\cal H}^{{\rm eff}} &=& -\frac{1}{ 2M_P^2} \mbox{\boldmath $\pi$}^2_a +
 M_P^2\left( - a^2 +
{4\Lambda\over 9\pi }~a^4
\right)  + \sum_{J,L}~{4\over 3\pi} m^2 a^2 \beta^{abNM}_{LJ}
{}~\beta^{abLJ}_{NM} \nonumber \\
&& + \sum_{\vert J-L\vert =1} \left[ \overline{\alpha}\pi~
\mbox{\boldmath
 $\pi$}_{\beta^{abLJ}_{NM}}
\mbox{\boldmath $\pi$}_{\beta^{abLJ}_{NM}} + \beta^{abNM}_{LJ}
\beta^{abLJ}_{NM}
 (L+J+1)^2\right]  + \mbox{\boldmath $\pi$}^2_{\chi} + {\overline{\alpha}\over
3\pi}
\left[
 \chi^2  - {3\pi\over
2\overline{\alpha}}\right]^2
\nonumber \\
&& + \sum_J \left\{
\overline{\alpha}\pi + \left[(-1)^{4J} \left({16\pi^2 J(J+1)\over
2J+1}\right)
 ~{3\pi
\over 4m^2}\right]~~
\left[ 1 + {1\over a(t)}\right] \right\}\mbox{\boldmath
 $\pi$}_{\beta^{abJJ}_{NM}}
\mbox{\boldmath $\pi$}_{\beta^{abJJ}_{NM}} \nonumber \\
&& + \sum_{J,L} {4\over \overline{\alpha}\pi}~\left[ 1 +
 \sqrt{{2\overline{\alpha}\over 3\pi}} \chi
\right]^2~~\beta^{abNM}_{LJ} ~\beta^{abLJ}_{NM} \ + 4\pi a^2 m^2~\left[ 1 +
\sqrt{{2\overline{\alpha}\over
3\pi}}\chi\right]^2,
\label{12}
\eea
where the canonical conjugate momenta of the dynamical variables are given by
\beq
\mbox{\boldmath $\pi$}_{a} = {\partial {\cal L}^{{\rm
eff}}\over\partial\dot a}
 = -\frac{\dot a}{N}~,~~
\mbox{\boldmath $\pi$}_{\chi} =
{\partial {\cal L}^{{\rm eff}}\over \partial \dot\chi} =
\frac{\dot\chi}{N}~\mbox{\boldmath $\pi$}_{\beta^{abNM}_{LJ}} = {\partial {\cal
L}^{{\rm
eff}}
 \over\partial\dot\beta^{abNM}_{LJ}}
= \frac{\dot\beta^{abNM}_{LJ}}{2N\pi\overline{\alpha}}~.
\label{13a}
\eeq

\beq
\mbox{\boldmath $\pi$}_{\beta^{abNM}_{JJ}} = {\partial {\cal L}^{{\rm
eff}}
 \over\partial\dot\beta^{abNM}_{JJ}}
= \frac{\dot\beta^{abNM}_{JJ}}{2N\pi\overline{\alpha}} -
\frac{1}{2\pi\overline{\alpha}}\frac{a}{N}
\alpha^{abJM}_{\phantom{^{JM}} N}
(-1)^{2J} \sqrt{{16\pi^2 J(J+1)\over 2J+1}},
\label{13b}
\eeq
${\cal L}^{{\rm eff}}$ denoting the effective Lagrangian density
arising from (\ref{1}) and the dots representing derivatives with respect to
the conformal time.

The Hamiltonian constraint, ${\cal H}^{{\rm eff}} = 0$, gives origin to the
Wheeler-DeWitt equation after promoting the canonical conjugate momenta
(\ref{13a}),(\ref{13b}) into operators:
\beq
\mbox{\boldmath $\pi$}_{a} = -i{\partial\over\partial
a}~,\mbox{\boldmath
 $\pi$}_{\chi} =
-i{\partial\over\partial
\chi}~,~~ \mbox{\boldmath $\pi$}_{\beta^{abNM}_{LJ(JJ)}} =
 -i{\partial\over\partial
\beta^{abLJ(JJ)}_{NM}}~,
{}~~\mbox{\boldmath $\pi$}_{a}^2 = - a^{-P}{\partial\over\partial a}~
\left( a^P{\partial\over\partial a}\right)~,
\label{14}
\eeq
\noindent
where in (\ref{14}) the last substitution parametrizes the operator order
ambiguity with $p$ being a real constant.
The Wheeler-DeWitt equation is obtained imposing that the Hamiltonian operator
annihilates the wave function
$\Psi [a,\beta^{abNM}_{LJ},\beta^{abNM}_{JJ},\chi]$.

A solution of the Wheeler-DeWitt equation
which corresponds to a classical behaviour
of its variables on some region of minisuperspace will have an
oscillatory WKB form as
\begin{equation}
\Psi_{(n)} [a,\mbox{\boldmath $A$}_\mu] = e^{iM_P^2 S_{(n)}(a)}
C_{(n)}(a)
 \psi_{(n)}
(a,\mbox{\boldmath $A$}_\mu)~.
\label{WKB}
\end{equation}

In order to make predictions concerning the behaviour of the scale factor, $a$,
for the U(1) case
one uses a coarse-grained description of the system
working out the reduced density matrix associated to (\ref{WKB})
\begin{equation}
\rho_R = \sum_{n,n'} e^{iM_P^2[S_{(n)}(a_1) - S_{(n')}(a_2)]}
C_{(n)}(a_1)C_{(n')}(a_2) {\cal I}_{n,n'}(a_2,a_1)~,
\label{red}
\end{equation}
where
\begin{equation}
{\cal I}_{n,n'}(a_2,a_1) \equiv  \int \psi^*_{(n')} (a_2,\mbox{\boldmath
 $A$}_{\mu}) \psi_{(n)}
(a_1,\mbox{\boldmath $A$}_{\mu}) d[\mbox{\boldmath $A$}_{\mu}]
= \Pi_J \int \psi^*_{J(n')} (a_2,\beta_J) \psi_{J(n)}
(a_1,\beta_J) [ d \beta_J]~.
\label{infl}
\end{equation}
The subindex $(n)$ labels the WKB branches. The term
${\cal I}_{n,n'}(a_2,a_1)$ contains
the environment influence on the system.

The decoherence process is successful when the non-diagonal terms $(n\neq n')$
in (\ref{red}) are vanishingly small. Hence, there will be no quantum
interference
between alternative histories if $ {\cal I}_{n,n'} \propto \delta_{n,n'}$.
Once that is achieved one can analyse the correlations in each classical
branch ($n = n'$). This can be performed looking at the reduced density matrix
or the to corresponding Wigner functional:
\begin{equation}
F_{W,(n)} (a,\pi_a) = \int_{-\infty}^{+\infty} d\Delta
[S'_{(n)}(a_1) S'_{(n)}(a_2)]^{-\frac{1}{2}} e^{-2i\pi_a\Delta}
e^{iM_P^2[S_{(n)}(a_1) - S_{(n)}(a_2)]} {\cal I}_{n,n}(a_2,a_1)
\label{wigf}
\end{equation}
where $\Delta = \frac{a_1 - a_2}{2}$.
A correlation among variables will correspond to a strong peak
about a classical trajectory in the phase space. The decoherence process
is strictly necessary to achieve that as the Wigner function associated to
(\ref{red}) does not have a single sharp peak even for a WKB Wigner function as
(\ref{wigf}); such a peak (and a clearly classical WKB evolution) is found
only among the $n=n'$ terms.
If the conditions to obtain an effective diagonalization of (\ref{wigf})
are met then the
interference between the different classical behaviours is {\em also}
highly suppressed.
Furthermore, ${\cal I}_{n,n'}(a_2,a_1)$ will be damped for
$|a_2 - a_1|\gg 1$ and the
reduced density matrix associated with (\ref{wigf}) will be diagonal
with respect to the variable $a$ \cite{OBPM}.

As far as the non-Abelian case is concerned,
correlations between each minisuperspace coordinate $a$ and $\chi$ and
their canonical momenta
can be analysed by examining peaks in the {\em reduced} Wigner function
\beq
F_{W1} (q^A,\mbox{\boldmath $\pi$}_{q_{1}}) =
\int d\mbox{\boldmath $\pi$}_{q_{2}} F_{W(n)}(q^A; \mbox{\boldmath
 $\pi$}_{q_{A}})
\label{redwigf}
\eeq
which is equivalent to
\beq
F_{W1} (q^A,\mbox{\boldmath $\pi$}_{q_{1}}) =
\exp \left[\epsilon^{11}\left[ \mbox{\boldmath $\pi$}_{q_{1}} - M_P^2
 \frac{\partial S}{\partial q^1} -
\tilde{\epsilon}^1\right]^2\right],
\label{redwigfa}
\eeq
where $\mbox{\boldmath $\pi$}_{q_{1}}$ is the momentum conjugate to
$q^A,A=1$.
The so-called strong correlation condition which ensures that the Wigner
function is sharply peaked, translates as
\beq
\epsilon^{AA} \ll \overline{\mbox{\boldmath $\pi$}_A}^2
\eeq
(where $\overline{\pi_A}$ is a typical value of the momentum along
the trajectory). On its hand, the so-called
strong decoherence condition which guarantees an effective diagonalization of
the reduced density matrix, consists in
\beq
\epsilon^{AA} \gg 1/q^A.
\eeq
Notice that in eqs. (\ref{redwigf}),
(\ref{redwigfa}) a sum over the $SO(3)$-group indexes is implicit
and that {\em all} modes have been taken into account.

As a summary (see Refs. [1,3,4] for
details),  we conclude that
a FRW  quantum cosmological model with   a massive vector field
and  global internal symmetry
possesses interesting properties in what concerns the decoherence process.
In the presence of a massive
vector field with U(1) global symmetry, the scale factor is the only decohered
quantity, while for the non-Abelian case with SO(3) global symmetry,
the scale factor and the homogeneous mode $\chi(t)$ are the expected decohered
variables.

As far as we restrict ourselves to expanding semiclassical solutions,
the models we propose
can be regarded to be essentially in the same footing
to the ones where decoherence of the metric degrees of freedom
is achieved via tracing out higher modes of self-interacting
scalar fields. In this respect, the inhomogeneous modes of massive vector
fields represent an interesting alternative to play the role
of environment.
However, interestingly, we find out that the strong decoherence
condition, for the $\chi$-field (parametrizing the non-Abelian
massive vector field homogeneous modes), is not fully satisfied [1].

\bigskip

{\large\bf Acknowledgments}

\bigskip
\indent
The authors gratefully acknowledge H.F. Dowker, B.L. Hu,
J.P. Paz and J. Mour\~ao for valuable comments.


\begin{thebibliography}{9}

\bibitem{OBPM} O. Bertolami and P.V. Moniz, ``Decoherence of
Friedmann-Robertson-Walker
Geometries in the Presence of Massive Vector Fields with U(1) or SO(3) Global
Symmetries", Preprint CERN-TH.7241/94,
DAMTP R-94/22; Bulletin Board  GR-QC 9410027, accepted for
publication in Nuclear Physics B.

\bibitem{PP} M.C. Bento, O. Bertolami, P.V. Moniz, J.M. Mour\~ao and P.M. S\'a,
\CQG {\bf 10} (1993) 285.


\bibitem{SS} O. Bertolami and P.V. Moniz, ``Decoherence of Homogeneous and
Isotropic Geometries in the Presence of Massive Vector Fields",
in Proceedings of the III National Meeting on Astronomy
and Astrophysics, July 1993, Lisbon, Portugal; Bulletin Board GR-QC 9407025.


\bibitem{SS1} O. Bertolami and P.V. Moniz,  ``Decoherence of Homogeneous and
Isotropic Metrics in the Presence of Massive Vector Fields",  Talk given at
7th Marcel Grossmann Meeting on General Relativity,
July 1994, Stanford, USA; Bulletin Board GR-QC-9409042.

\bibitem{LLL} O. Bertolami and J.M. Mour\~ao, \CQG {\bf 8} (1991) 1271.


\end{thebibliography}
\end{document}